# Traceable Trust for action-ready artificial intelligence in bioscience

Huayu Xin[1*], Yizhi Cai[2], Mukilan Deivarajan Suresh[3,4,5], Gavin Michael Farrell[6,7], Iwona Gajda[8], Charlie Harrison[9,10], Conor Houghton[11], Mato Lagator[12], Yang Lu[13], Virginia Portillo[14], Reyer Zwiggelaar[9*] & Sebastian Lobentanzer[15,16,17,18*]

[1] Science, Technology and Innovation Studies, School of Social and Political Science, University of Edinburgh, Edinburgh, UK.
[2] Manchester Institute of Biotechnology, University of Manchester, Manchester, UK.
[3] School of Natural and Environmental Sciences, Newcastle University, Newcastle upon Tyne, UK.
[4] School of Computing, Newcastle University, Newcastle upon Tyne, UK.
[5] Protecting Crops and Environment, Rothamsted Research, Harpenden, UK.
[6] Department of Biomedical Sciences, University of Padova, Padova, Italy.
[7] European Molecular Biology Laboratory, European Bioinformatics Institute (EMBL-EBI), Hinxton, UK.
[8] Bristol Robotics Laboratory, University of the West of England, Bristol, UK.
[9] Department of Computer Science, Aberystwyth University, Aberystwyth, UK.
[10] Informatics Research Institute, University of Louisiana at Lafayette, Lafayette, LA, USA.
[11] School of Engineering Mathematics and Technology, University of Bristol, Bristol, UK.
[12] Division of Evolution, Infection and Genomic Sciences, School of Biological Sciences, Faculty of Biology, Medicine and Health, University of Manchester, Manchester, UK.
[13] Department of Computer Science, School of Science, Loughborough University, Loughborough, UK.
[14] School of Computer Science, University of Nottingham, Nottingham, UK.
[15] Helmholtz AI, Artificial Intelligence Cooperation Unit of the Helmholtz Association, Germany.
[16] Institute of Computational Biology, Computational Health Center, Helmholtz Center, Munich, Germany.
[17] School of Computation, Information and Technology, Technical University of Munich, Germany.
[18] German Center for Diabetes Research, Munich, Germany.

***Correspondence to:** Huayu Xin (huayu.xin@ed.ac.uk), Reyer Zwiggelaar (rrz@aber.ac.uk) and Sebastian Lobentanzer (sebastian.lobentanzer@helmholtz-munich.de).

**Abstract**

Artificial intelligence (AI) is becoming part of the working infrastructure of the biosciences. AI models can predict biomolecular structures, design proteins, rank variants, annotate images, recommend strains and optimise experimental conditions. We argue that the decision to use an AI output to guide laboratory action is a key juncture for trustworthy research and should follow a defined, reviewable process. We propose Traceable Trust as a proportionate assessment-and-design framework for this output-to-action boundary. It asks what evidence supports the output, what capability is being claimed, what agency has been delegated, what threshold authorises action, who can override it and how outcomes inform later decisions. We illustrate the framework through three case studies spanning ecosystem resources, project design and laboratory action. Together, the cases show how trust can be documented where AI outputs begin to shape scientific work.

## 1. Assessing trust at the output-to-action boundary

Artificial intelligence (AI) systems can now shape which biological experiments are prioritised and how laboratory resources are allocated. Structure-prediction algorithms can inform hypotheses about molecular interactions; generative protein-design models can produce candidates for synthesis; and recommendation and optimisation workflows can help select strains or media for subsequent development[1–5]. These systems range from specialised machine-learning models to agentic systems with some capacity to plan, recommend or coordinate sequences of laboratory tasks. Their outputs can shape time, reagents, samples, instrument queues and the attention of research teams. Without a reviewable process, outputs may be followed, rejected or modified, leaving no record to support later reconstruction or audit.

We argue that, in experimental bioscience, trust in AI systems is usefully assessed at the output-to-action boundary. This boundary lies between an AI output (such as a suggestion, prediction, ranking or plan) and its implementation in a research workflow. The practical question is whether a particular output is supported by sufficient evidence and checks to justify a particular action: repeating a measurement, ordering variants, committing a liquid-handler run or changing the next design-build-test-learn (DBTL) round.

Trust therefore has a practical dimension alongside attitudes towards AI systems. As AI recommendations increasingly shape experimental decisions, researchers need to know what level of reliance an output can support. That judgement depends on model performance, the available evidence, the context of use, the actors involved, the required checks and the anticipated and recorded consequences of the action.

In this context, trust refers to a user's willingness to rely on an AI output in a specific setting, while trustworthiness concerns whether that reliance is warranted. The trustworthy AI literature commonly distinguishes these concepts and emphasises the importance of context and the risks of unmonitored reliance[6–9].

This framing also draws on science and technology studies, where agency may be understood as emerging from relations among people, models, instruments and records[10,11]. In laboratory AI, a transition record can make these relations reviewable by documenting roles, thresholds, available evidence, authorisation and how failed or negative outcomes shape subsequent decisions.

The bioscience community has made substantial progress in reporting and documenting experimental procedures and outcomes. The Data, Optimization, Model and Evaluation (DOME) recommendations set expectations for supervised machine-learning validation in biology, and the DOME Registry supports their implementation[12,13]. The Recommendations for Machine-learning-based Science (REFORMS) checklist provides broader guidance across disciplines[14]. Datasheets and Model Cards provide templates for data and model documentation[15,16]. The Findable, Accessible, Interoperable and Reusable (FAIR) principles, Research Object Crate (RO-Crate) and Workflow Run RO-Crate support reusable digital research artefacts and computational workflow provenance[17–19]. AI-Ready Bioscience Datasets (AIRBDS) Core complements these efforts with a quantitative, evidence-based checklist for assessing the AI-readiness of bioscience datasets[20]. Work on leakage and reproducibility in machine-learning-based science further shows why disclosure matters[21]. Traceable Trust concentrates on a complementary practical question: what is needed to support the trustworthy and effective use of AI outputs in laboratory decision-making?

The distinction between AI-readiness and action readiness helps locate this gap. AI-readiness concerns whether data, metadata, software and infrastructure are prepared for effective AI use[22]. Action readiness concerns whether a particular AI output can be justified as the basis for a research action. The two can diverge. A strong model can still sit within a weak workflow: a high-scoring prediction may fall outside its validated domain, be insufficiently calibrated, enter an automation loop with excessive permissions, or be carried forward without complete records of failed and negative outcomes.

Here we introduce Traceable Trust as a framework for assessing and designing the transition from AI output to laboratory action. The aim is a proportionate record with enough evidence for another competent researcher to reconstruct, challenge and revise the decision. Traceable Trust complements wider risk and deployment frameworks, including the US National Institute of Standards and Technology (NIST) AI Risk Management Framework and clinical AI reporting and deployment guidance. Its focus is the transition at which an AI output is allowed to influence experimental work[23–25].

Traceable Trust therefore covers both sides of the trust–trustworthiness distinction in experimental bioscience. Its six components specify the evidence needed to justify a particular output-to-action decision, and the outcome record supports subsequent review when results are poor or unexpected.

## 2. Traceable Trust as an assessment-and-design framework

Traceable Trust is designed to assess whether an AI output can support a defined experimental action. The assessment uses six elements (Table 1): (1) evidential provenance; (2) a situated capability claim; (3) a delegated agency boundary; (4) an action threshold with safeguards; (5) an accountable authorisation and override route; and (6) an outcome, exception and revision record. Fig. 1 summarises the transition and case layers.

**Table 1 | Six components of Traceable Trust**

| Component | Assessment question | Minimum evidence |
|---|---|---|
| Evidential provenance | What evidence produced this output? | Data source; sample, plate or batch identity or pseudonym; preprocessing; model and workflow version; prompt or configuration; timestamp; domain context; known exclusions. |
| Situated capability claim | What has been validated in this setting? | Validated task; intended use; domain boundaries; comparator or baseline; performance range; calibration or uncertainty; known failure modes. |
| Delegated agency boundary | What may the system influence? | Agency level; permissions; permitted and prohibited actions; data access; resource limits; out-of-scope actions. |
| Action threshold and safeguards | When is the output sufficient for action? | Predefined numerical or procedural threshold; quality-control rules; replicate and control requirements; uncertainty limits; stopping rules; escalation triggers. |

| Component | Assessment question | Minimum evidence |
|---|---|---|
| Accountable authorisation and override | Who can accept, block or change the action? | Named role; approval route; override log; escalation path; responsibility for review. |
| Outcome, exception and revision record | What happened, and how does it affect the next decision? | Raw results; invalid, negative and failed outcomes; deviations; human overrides; model or workflow update; rationale for the next round. |

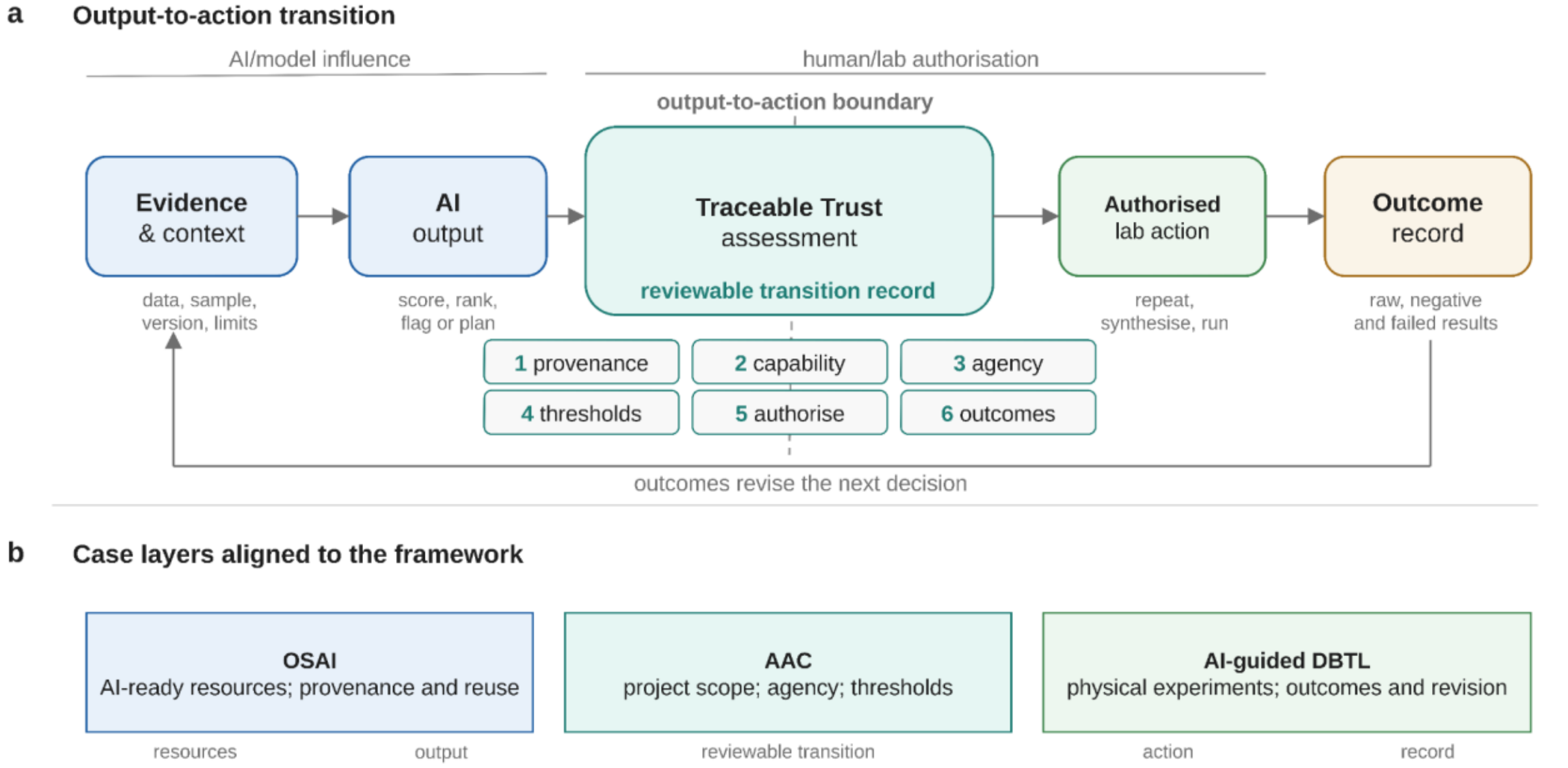

**Fig. 1 | Output-to-action transition.** a, The transition record marks the review point between AI output and authorised laboratory work. b, The cases locate this transition across resource, project-design and experimental-action layers. OSAI, open and sustainable AI; AAC, Agentic Automation Canvas; DBTL, design-build-test-learn.

The six components form a compact set for the point at which a laboratory is considering action on an AI output. Benefit, data sensitivity, environmental cost and institutional risk, though not separate components of the framework, remain relevant contextual variables that shape thresholds, safeguards and authorisation. This arrangement keeps the transition record proportionate.

Traceable Trust then asks two practical questions. What has the system been validated to do in this setting? This concerns *capability*: classification, ranking, generation, optimisation or planning within a specified domain. What may the laboratory allow the system to influence? This concerns *agency*. The same model can serve as a read-only adviser, a copilot, a planner, a workflow trigger or an administrative system that commits resources. Judgements about appropriate use therefore depend on both capability and the permissions built into the surrounding workflow.

For practical assessment, delegated agency can be described in four levels. Advisory use produces suggestions for people to review. Coordinative use organises tasks, such as drafting a plate map or reagent list. Executive use initiates software steps or instrument runs under defined constraints, usually after human approval. Resource-committing use consumes samples, reagents or instrument time, or shifts the next research direction. Advisory use may require less evidence than executive or resource-committing use, which will often call for stronger validation and explicit approval. Thresholds and safeguards should reflect the application and delegated agency, drawing where relevant on evidence about robustness, uncertainty calibration, reproducibility, explainability, privacy and fairness.

Building on work on responsible AI[26] and explainable AI[27], we argue that action-ready explainability supports this judgement by helping researchers understand why a system produced a particular output, what evidence it used, which uncertainties remain and whether the output lies within the validated domain. Traceable Trust therefore asks for explanations that are useful at the point of action. For example, a quantitative polymerase chain reaction (qPCR) repeat flag should point to the file, well, rule or model signal that produced it; a DBTL recommendation should record whether a candidate was chosen for predicted yield, uncertainty reduction, controls or replication. For agentic systems, this action-level explanation can make outputs easier to reproduce, contest and revise.

The Traceable Trust transition record also makes these relations visible in practice. It records how AI outputs, data, instruments, roles, thresholds and authorisation routes come together and how events such as instrument errors, borderline amplification curves, contaminated wells or human overrides interrupt or revise that configuration.

## 3. Case studies

We provide three case studies, which trace different parts of the path from reusable AI resources to laboratory action. Open and sustainable AI (OSAI) operates mainly at the ecosystem level, the Agentic Automation Canvas (AAC) at the project-design level, and AI-guided DBTL laboratories at the experimental-action level. The comparison shows what each approach contributes to Traceable Trust and where a transition record adds information.

### 3.1 Open and Sustainable AI and ecosystem-level traceability

OSAI is a set of principles and resources for developing and sharing AI in the life sciences so that models and workflows are easier to find, reuse, reproduce and maintain. It is a useful case study because it addresses the AI ecosystem: the resource base from which local laboratory decisions draw. The OSAI Perspective argues that poor reusability and reproducibility can erode trust in life-science AI and that these problems also affect environmental sustainability. Its practical contribution is a set of recommendations mapped to more than 300 ecosystem components, with guiding implementation pathways[28].

Suppose a group plans to reuse an open protein-design or enzyme-variant prioritisation model before ordering variants for synthesis. The group would need to identify the model release, code, licence, training-data description, evaluation evidence, archive status and practical guidance for reuse. OSAI helps at this selection stage by pointing researchers to open components, repositories, standards and implementation routes.

In Traceable Trust terms, OSAI contributes clearly to evidential provenance. It helps researchers check whether the resources behind an AI workflow can be found, reused, compared and maintained. It also supports AI-readiness through links with research data management toolkits, Croissant metadata for machine-learning-ready datasets[29,30], model openness frameworks and open implementations such as OpenFold[31,32].

OSAI mainly provides resource-level evidence. Local teams add a situated capability claim, delegated agency boundary, action threshold and safeguards, authorisation route, and outcome record for each specific action. This division clarifies the scale at which OSAI works: it improves the resource base from which local action decisions can be made.

One possible extension for OSAI-like infrastructures is transition-oriented metadata. Components could be tagged by the transition they support: data-to-model, model-to-output, output-to-action or outcome-to-revision. Component pages could also state whether they support versioned export, what local evidence is needed for a capability claim, and how they connect to AAC canvases, electronic laboratory notebooks (ELNs), laboratory information management systems (LIMS) or RO-Crate packages. This could make ecosystem resources easier to use in action-level assessments while retaining OSAI's ecosystem focus.

### 3.2 Agentic Automation Canvas and agency boundary

AAC is a structured framework for designing agentic AI projects before a system is built or deployed. It asks project teams to describe the proposed automation project, including its scope, user expectations, benefit metrics, feasibility, governance stages, data access and sensitivity, and intended outcomes. It is implemented as a semantic-web-compatible metadata schema mapped to

Schema.org and the World Wide Web Consortium (W3C) Data Catalog Vocabulary (DCAT), with client-side validation and export as RO-Crates[33].

A qPCR example illustrates how AAC connects to Traceable Trust. A molecular biology laboratory processes plates each day. A proposed AI system reads instrument export files, a plate map and experiment records; reviews amplification curves, quantification cycle (Cq) values, controls and, for dye-based assays, melt curves; and flags wells that may need repeat measurement. Its permitted role is review support: the system drafts a repeat list and reagent estimate, while human operators retain control of automated runs and reported results. Existing qPCR reporting standards are relevant because the original Minimum Information for Publication of Quantitative Real-Time PCR Experiments (MIQE) guidelines and the updated MIQE 2.0 guidelines both emphasise sufficient experimental detail, assay characteristics, analysis information and transparent reporting[34,35].

Using AAC for this project would make the planned role of the AI system explicit. The project scope would define recommendations for human review. Benefit and feasibility fields would record expected review-time savings, an acceptable false-positive burden, the minimum reduction in missed control failures, required inputs and known instrument-specific limits. The governance field would define staged validation, from advisory flagging to draft repeat lists after local validation. The data-sensitivity and outcomes fields would require pseudonymised identifiers and record accepted, rejected or overturned recommendations.

Within the Traceable Trust framework, AAC contributes directly to the situated capability claim, delegated agency boundary, action threshold and accountable authorisation route. It makes the planned role of an AI system visible before the system is connected to the laboratory. During use, AAC would need to be linked to evidence generated in the workflow. A canvas is a design record; ELN or LIMS entries, instrument files, human overrides, repeat outcomes and failures provide the action-level record for each plate. Policy Cards provide machine-readable runtime constraints for autonomous agents, while Traceable Trust links such constraints to provenance, situated capability, human authorisation and experimental outcomes at the scientific output-to-action transition. Workflow Run RO-Crate can provide an adjacent execution-provenance layer when a project moves from planning to operation[19,33,36].

The AAC case also shows how Traceable Trust makes transparency action-specific. For a repeat decision, the laboratory mainly needs a record of which files were read, what rule or model signal produced the flag, what action was allowed, who reviewed it and what happened after the repeat. This trace provides the information needed to review the repeat decision.

### 3.3 AI-guided DBTL laboratories and action-level traceability

The third case is AI-guided DBTL in bioengineering. It draws on a small set of illustrative systems. BioAutomata is a robotic and machine-learning platform for biosystems design: in lycopene pathway optimisation, a paired predictive model and Bayesian algorithm selected experiments for the Illinois Biological Foundry for Advanced Biomanufacturing (iBioFAB). The platform evaluated less than 1% of the possible variants and outperformed random screening by 77% (ref. 37). The Automated Recommendation Tool (ART) shows the recommendation layer: it uses machine learning and probabilistic modelling to recommend strains for the next engineering cycle and to provide probabilistic predictions of production levels[4]. Machine-learning-guided Experimental Trials for Improvement of Systems (METIS) illustrates active learning under wet-lab constraints. It was applied to cell-free transcription and translation, genetic circuits and a 27-variable synthetic $CO_2$-fixation cycle, exploring a space described as $10^{25}$ possible conditions with 1,000 experiments (ref. 5).

Together, these studies illustrate the core laboratory issue: in DBTL workflows, a model-guided learn step can shape which physical work happens next. BioAutomata anchors physical execution, ART the learn-step recommendation and METIS data-efficient active learning. Recent work on AI-native autonomous biomolecular laboratories points towards multi-user, multi-instrument systems, and the same issue is already visible in these DBTL studies[38]. Automated DBTL pipelines[39] and biofoundry networks[40] may make the issue more consequential because recommendations can draw on shared instruments, standardised workflows and scarce technical capacity.

Consider a laboratory optimising enzyme variants or pathway conditions in a 96-well format. After an initial round, the model ranks the next set of variants or media conditions. Some options have high predicted yield, some reduce uncertainty, and others serve as controls or replicates. The resulting work may involve assembling constructs, inoculating cultures, running assays or reserving an automated liquid handler. To justify reliance on the recommendation, the team needs to explain why the batch was selected, why other candidates were postponed, what uncertainty was accepted and which quality-control rules had to be met before the run.

In such a setting, a Traceable Trust assessment could record one transition per experimental round or batch. The record could include the training-data snapshot, model version, acquisition strategy, calibration or uncertainty estimate, plate layout, replicate and control requirements, failed or excluded candidates, approval route and stopping conditions. The outcome record could retain raw assay files, invalid-well calls, contamination events, sequencing confirmation where relevant, low-yield and negative results, and human overrides. In active-learning settings, low-yield results can be as informative as high-yield results[5]; discarding them may distort the landscape being learnt.

This case aligns closely with the action-threshold and outcome-record components of Traceable Trust. Implementation may be demanding. The illustrative optimisation studies report overall performance more fully than candidate-level authorisation. A lightweight transition record could link the model recommendation, wet-lab schedule, human approval, instrument output and revision of the next model. Workflow Run RO-Crate can package computational workflow provenance, while laboratory action records add sample, plate, reagent and instrument context[19].

## 4. Aligning the case studies

The three cases locate Traceable Trust at complementary layers. OSAI supports AI-ready resources; AAC defines project scope and agency before deployment; and AI-guided DBTL laboratories show how recommendations may select new physical experiments. Together, they trace the path from reusable resources to action-level evidence. Table 2 summarises how each case contributes to the six components.

**Table 2 | How the three case studies contribute to Traceable Trust**

| Traceable Trust component | OSAI | AAC | AI-guided DBTL laboratory |
|---|---|---|---|
| Evidential provenance | Provides discoverable resources, versions, licences, repositories and reusable components. | Records planned inputs and exports; links to plate-level files strengthen use. | Requires snapshots of training data, model versions, acquisition strategy, plate layout and instrument context. |
| Situated capability claim | Supports reuse decisions; local laboratories validate capability in context. | Defines intended task, benefits, limits and staged validation before deployment. | Uses round- or batch-level records of performance, uncertainty and failure modes. |
| Delegated agency boundary | Provides resource-level evidence; local teams define what the system may do. | Specifies whether the AI system advises, coordinates, executes or commits resources. | Shows how recommendations can shape physical experiments and instrument use; records make this boundary visible. |

| **Traceable Trust component** | **OSAI** | **AAC** | **AI-guided DBTL laboratory** |
|---|---|---|---|
| Action threshold and safeguards | Provides supporting evidence while local teams set laboratory thresholds. | Defines planned thresholds, quality-control rules, escalation and approval conditions. | Uses thresholds to select candidates, controls, replicates and uncertain options for the next batch. |
| Accountable authorisation and override | Provides resource evidence while local teams authorise specific wet-lab actions. | Defines design-stage approval; runtime records show who approved, rejected or changed recommendations. | Records approval for each round, especially where automated platforms consume scarce resources. |
| Outcome, exception and revision record | Supports export links that connect resources to later records. | Links to ELN, LIMS, instrument logs and repeat outcomes make outcome records usable. | Uses failed, invalid, negative and low-yield results to inform the next model and decision. |

The comparison suggests that a reviewable output-to-action transition draws on evidence from all three layers: traceable resources, roles and thresholds defined before action, and outcomes that inform the next decision.

## 5. Evaluating Traceable Trust

Trustworthy AI assurance approaches assemble evidence about whether an AI system meets legal, ethical and technical criteria across its lifecycle, with limited focus on relational assurance strategies[41]. Traceable Trust adapts evidence-based trustworthy AI assurance to a narrower scale, focussing on the output-to-action boundary.

We propose six practical indicators that map across the six components. Some test a specific component, while others assess how the transition record works as a whole. Provenance completeness asks whether each action records the relevant evidence, model or workflow version, threshold and approval. Decision reconstruction time asks how quickly a team can reconstruct why a particular action was taken. Threshold compliance asks whether a run met the predefined quality-control, replicate and uncertainty requirements. Override learning asks whether human rejections and changes are recorded and used to improve the workflow. Failure capture asks whether failed, invalid and negative outcomes receive the same attention as successful results. User burden asks whether the added record-keeping is manageable in routine laboratory work.

The indicators should be applied proportionately to the setting. A qPCR repeat recommender may need only a simple record of flags, reviews, repeats and outcomes. A protein-design or DBTL platform will usually require richer information about model versions, candidate generation, batch design, instrument logs and failed constructs. The aim is enough relevant, credible information to review an action while keeping routine work manageable.

The components provide a common set of questions. Their evidential depth and practical weight will vary with the action, risk and user. A partial record can expose missing evidence and support a correspondingly limited claim about action readiness. Quantitative measures and qualitative review may both be useful, depending on the setting. The indicators are diagnostic prompts for discussion; compliance scores, where used, should remain secondary to decision reconstruction.

**6. Implications for journals, research institutions and infrastructures**

Traceable Trust provides a basis for coordination across journals, laboratories, research institutions and infrastructures that share responsibility for scientific integrity and critical judgement as AI moves from prediction to physical action.

For journals, Traceable Trust complements reporting standards. Methods sections and supplementary information can specify which AI outputs changed experimental actions and report the relevant provenance, capability claim, agency level, threshold and outcome record. DOME, REFORMS and image-analysis checklists show how communities can translate complex methodological concerns into reportable expectations[12,14,42]. One proportionate option is to ask authors to identify the output-to-action transitions that materially affected the reported work. Claims about action readiness should align with the available evidence, and authors should state limitations clearly[43].

For laboratories, the framework can guide the adoption of agentic tools and automation. A practical route is to classify the delegated agency level, define action thresholds before the first run and connect outcome records to the next model or workflow decision. Initial use can focus on high-value transitions such as repeat recommendations, variant selection, microscopy triage, plate scheduling or automated protocol changes. The aim is to make each AI-guided action reviewable before the system becomes part of routine work.

For research institutions and universities, deploying agentic AI within specialised research domains depends on both training and domain-specific governance. Early-career scientists need opportunities to question AI outputs, explain when and why they rely on them, and record how evidence, uncertainty and outcomes shaped a decision. Embedding Traceable Trust in supervision, methods teaching and research-integrity training can support critical thinking and reflective practice across the workflow, including in human-in-the-loop decisions. Such governance should involve subject specialists who understand both the opportunities and limitations of AI in their fields. Institutions also need to develop AI literacy, promote responsible use, safeguard sensitive data and protect academic integrity. Traceable Trust connects these responsibilities to reviewable evidence, permissions, authorisation and outcomes[44,45].

For research infrastructures, Traceable Trust bridges AI-ready and action-ready resources. OSAI-like catalogues help users find components. AAC-like canvases define the planned role of those components. RO-Crate and Workflow Run RO-Crate package computational execution records. ELNs, LIMS and instrument systems carry the corresponding sample, plate and wet-lab context.

The practical challenge is interoperability: records need to move across these systems while minimising duplicate data entry.

Similar questions arise beyond bioscience wherever AI outputs allocate material resources or trigger physical action, including in chemistry, materials science, engineering and environmental research. They also arise in participant-facing research. In biobanking, generative AI may help participants understand how donated samples and data are used[46]. Yet using technology to manage consent does not necessarily build trust and may instead reduce it when autonomy, social relationships or institutional responsibility are obscured[47]. A verifiable trust architecture therefore needs transparent records, accountable governance and routes for participants to question or contest data use. Thresholds and records vary by field, but the core questions about evidence, delegated agency, authorisation and outcomes remain.

For science and technology studies and responsible innovation research, Traceable Trust offers an empirical object for studying the practical work through which an action-ready arrangement is assembled, authorised, interrupted and revised[11]. It invites studies of how delegated agency is negotiated, how authority is distributed between bench scientists and automation teams, which failures are preserved, and whether trace records support scientific judgement or become paperwork. Such studies can help keep the framework from becoming a narrow technical checklist. AI systems can support analysis and decision-making but may also encourage users to overestimate their understanding of complex processes[48]. This risk highlights the need for critical evaluation and human oversight. Trust in AI depends not only on confidence in the model but also on confidence in the transparency, validation and governance processes surrounding its use[49].

## 7. Limitations and next steps

This Perspective proposes a framework that still requires empirical validation. A comparative metric for traceability practices across laboratories also remains to be developed. Validation should include laboratories with different levels of automation, data sensitivity, staff capacity and regulatory exposure, with close attention to burden because excessive form-filling can undermine scientific practice. Deliberate misuse and dual-use oversight fall beyond the scope of this perspective and merit separate treatment in AI and biosecurity research. Related work may still draw on similar output-to-action records.

Several research questions follow. Which transition records can be generated automatically from existing ELN, LIMS and instrument logs? Which action thresholds are meaningful for different tasks, such as qPCR repeats, variant synthesis, microscopy triage or DBTL optimisation? How might journals treat AI-generated recommendations that change experimental allocation before the final result is available? Which failures need to be retained or reported to make later model updates scientifically trustworthy?

The case selection also has limits. OSAI is an ecosystem-level case, AAC a design-stage case and DBTL a literature-derived laboratory case. Future work could compare lightweight transition records with existing documentation and test an openly available implementation across bioscience and other scientific and engineering fields. It could also develop a stewardship model for updating and verifying the framework with researchers, journals, institutions and infrastructure providers. Further questions include who benefits from traceability, who carries the recording burden and who has authority to reinterpret a trace when something goes wrong. These limits also support an anticipatory approach to governance. As AI systems begin to shape experimental action before their institutional forms are settled, laboratories and institutions may benefit from considering responsibilities, burdens and failure modes before traceability becomes an after-the-fact audit trail[50,51].

## 8. Conclusion

Bioscience has made progress in AI performance, reporting and openness. A pressing trust question concerns the conditions under which an AI output can guide a real experimental step. Traceable Trust offers researchers, journals, institutions and infrastructures a common language for those conditions. It asks what evidence produced the output, what capability is being claimed in context, what agency has been delegated, what threshold permits action, who can approve or stop the action, and how the outcome informs subsequent decisions. Making this transition visible can support challenge, reconstruction and revision. In this form, the evidential and decision-making process that supports trust can become a reviewable part of scientific work.

**Acknowledgements**

We thank AIBIO-UK and its Responsible Research Working Group for supporting and facilitating this work, and Maria Temenou, Network Manager for AIBIO-UK, for her support with coordination.

**Funding**

Mukilan Deivarajan Suresh received funding from the BBSRC NLD DTP (BB/T008695/1). Gavin Michael Farrell is funded by ELIXIR-STEERS (101131096), supported through Silvio Tosatto's BioComputingUP lab, University of Padova. Charlie Harrison and Reyer Zwiggelaar are supported by BBSRC through AIBIO-UK (BB/Y006933/1). Virginia Portillo is supported by Responsible AI UK (EP/Y009800/1).

**Author Contributions**

Conceptualization: Huayu Xin, Sebastian Lobentanzer. Supervision: Sebastian Lobentanzer, Reyer Zwiggelaar. Project administration: Reyer Zwiggelaar, Charlie Harrison. Investigation: Mato Lagator, Gavin Michael Farrell, Sebastian Lobentanzer. Writing – original draft: Huayu Xin. Writing – review & editing: all authors.

**Competing interests**

The authors declare no competing interests.